\documentclass[5p,times,twocolumn]{elsarticle}
\usepackage{graphicx}
\usepackage{amsmath}
\usepackage{listings}
\begin{document}

\begin{frontmatter}


\title{Numerical prediction of the piezoelectric transducer response in the
acoustic nearfield using a one-dimensional electromechanical finite difference
approach}

\author{O.\ Melchert} \ead{oliver.melchert@hot.uni-hannover.de}
\author{E.\, Blumenr\"other}
\author{M.\, Wollweber}
\author{B.\, Roth}
%
\address{Hannover Centre for Optical Technologies (HOT), 
Leibniz Universit\"at Hannover, Nienburger Str.\,17, D-30167 Hannover, Germany}

\begin{abstract}
We present a simple electromechanical finite difference model to study the
response of a piezoelectric polyvinylidenflourid (PVDF) transducer to
optoacoustic (OA) pressure waves in the acoustic nearfield prior to thermal
relaxation of the OA source volume. The assumption of nearfield conditions,
i.e.\ the absence of acoustic diffraction, allows to treat the problem using a
one-dimensional numerical approach.  Therein, the computational domain is
modeled as an inhomogeneous elastic medium, characterized by its local wave
velocities and densities, allowing to explore the effect of stepwise impedance
changes on the stress wave propagation.  The transducer is modeled as a thin
piezoelectric ``sensing'' layer and the electromechanical coupling is
accomplished by means of the respective linear constituting equations.
Considering a low-pass characteristic of the full experimental setup, we obtain
the resulting transducer signal.  Complementing transducer signals measured in
a controlled laboratory experiment with numerical simulations that result from
a model of the experimental setup, we find that, bearing in mind the apparent
limitations of the one-dimensional approach, the simulated transducer signals
can be used very well to predict and interpret the experimental findings.
\end{abstract}

\begin{keyword}
optoacoustics \sep piecewise homogeneous elastic media \sep finite difference
model \sep piezoelectric transducer \sep Python
\PACS 02.60.Cb \sep 43.35.+d \sep 77.65.Ly 
\end{keyword}

\end{frontmatter}

\section{Introduction}

Optoacoustics (OAs) can be considered a compound phenomenon, consisting of two
distinct processes that occur on different time-scales: fast optical absorption
of laser energy inducing a photothermal heating of the absorbing media, and,
subsequently, the emission of comparatively slow acoustic stress waves due to
thermoelastic expansion and stress field relaxation
\cite{Sigrist:1978,Scruby:1980,Sigrist:1986,Tam:1986,Kruger:1995}.  Albeit
thermoelastic expansion succeeds the absorption of laser light, it is by no
means the only energy conversion process that supports the production of
optoacoustic signals \cite{Hutchins:1986,Tam:1986,Sigrist:1986}. E.g., in case
of laser generation of stress waves in liquid one might identify three relevant
mechanisms, i.e.\ thermoelastic expansion, vaporization, and dielectric
breakdown, occurring at increasing deposited power-densities
\cite{Hutchins:1986}.  Owing to the possible generation and propagation of
transverse vibrations, the production of acoustic waves in solid is somewhat
more intricate \cite{Scruby:1980}.  However, in the absence of any change in
state of the underlying medium, thermoelastic expansion due to the absorption
of laser energy can be considered the dominant conversion mechanism
\cite{Hutchins:1986,Tam:1986}.  Compared to the propagation of acoustic stress
waves, which proceeds on a microsecond timescale, the optical absorption is
assumed to occur instantaneously.  Assuming the optical absorption to be
instantaneous has consequences for the theoretical treatment of the problem
inherent dynamics \cite{Diebold:1990,Diebold:1991,Kruger:1995}. It not only
allows to decouple the optical absorption problem from the acoustic propagation
problem but also allows to simplify the latter as discussed in the remainder.

Here, we present a combined study, complementing measurements on a controlled
experimental setup with custom numerical simulations in terms of a finite
difference model of the underlying physical processes.  In contrast to recent
studies wherein we discussed measurement, simulation and approximate inversion
of OA signals observed for layered PVA-H (polyvinyl alcohol hydrogel) phantoms
in the acoustic farfield \cite{Blumenroether:2016,Melchert:2017}, the specific
object of the presented study is to model the observed PVDF
(polyvinylidenfluorid) transducer response resulting from the subtleties of the
source volume in the acoustic nearfield.  The acoustic properties of the source
volume, determining the propagation of the stress waves and their behavior upon
crossing inter-layer boundaries, are considered to be piecewise homogeneous and
not too rapidly varying (as, e.g., opposed to the scenario considered by Refs.\
\cite{Santosa:1991,Fogarty:1999}).
If the experimental setup satisfies nearfield conditions, i.e.\ under the
assumption of plane acoustic waves, and for a translational symmetry of the
region of interest in the plane perpendicular to the direction of the
propagating stress waves, the evolution of the stress profile within the domain
can be modeled in terms of the equations of one-dimensional ($1$D) linear
elasticity \cite{Landau:1975,Tam:1986,Irgens:2008}. The response of a
piezoelectric sensing layer to trespassing stress waves is then included by the
electromechanical coupling to the constituting equations of linear
piezoelectricity \cite{Szabo:2004}. 
The article presents an application for a simple and efficient effectively $1$D
approach for the solution of a $3$D problem. Note that quite similar $1$D
approaches where considered in the literature to study, e.g., complex
transients in pulsed OA spectroscopy \cite{Schurig:1993}, surface heating and
energy transfer in pulsed microwave catalytic systems \cite{Wan:1994}, and,
thin-film piezoelectric ultrasonic sensors \cite{Gonzales:2014}.

The article is organized as follows. In section \ref{sec:S2} we discuss the
computational model for the numerical simulation of stress wave sensing in
piecewise homogeneous elastic media in detail. Therefore, we discuss the finite
difference stencils used to approximate the underlying differential equations
of the continuous models (i.e.\ linear elasticity and linear piezoelectricity)
and illustrate their implementation in terms of {\tt python}
modules\footnote{For conciseness, the presented code listings are not
documented thoroughly. In general, this has to be considered bad programming
style.  However, note that the code in the supplementary material under Ref.\
\cite{Melchert_gitHub_1DFD:2017} is documented well.} that are later used for
the numerical experiments discussed in section \ref{sec:S4}. In section
\ref{sec:S3} we point out limitations of the presented approach and conclude
with a summary in section \ref{sec:S5}.

\section{Modeling stress wave sensing in a piecewise homogeneous elastic medium}
\label{sec:S2}

In our one-dimensional approach, the computational domain represents a
heterogeneous elastic medium for which the acoustic displacement $r(z,t)$ at the
field point $z$ defines the material velocity field via $u(z,t) =\dot{r}(z,t)$.
Material velocity $u(z,t)$ and excess pressure $p(z,t)$ are related via the set
of coupled first-order differential equations of linear elasticity
\cite{Landau:1975,Irgens:2008}
\begin{subequations}
 \begin{align}
    K(z)\,\partial_z u(z,t) + \partial_t p(z,t) = 0, \label{eq:linEl1}\\
    \rho(z)\,\partial_t u(z,t) + \partial_z p(z,t) = 0. \label{eq:linEl2}
  \end{align}
\end{subequations}
Therein $\rho(z)$ denotes the local density and 
$K(z)\equiv \rho(z) c(z)^2$ signifies the bulk modulus of elasticity wherein
$c(z)$ refers to the local speed of sound. 

Subsequently we discuss the finite-difference approach used to model the
piezoelectric ``sensing'' of acoustic stress waves. In subsection
\ref{ssect:domain} we introduce the data structure used to represent the
piecewise homogeneous medium. In subsection \ref{ssect:opt} we elaborate on the
mechanism of optoacoustic signal generation, responsible for the generation of
initial stress profiles within our model. The finite difference approach for
the propagation of the acoustic stress waves in piecewise homogeneous elastic
media is illustrated in subsection \ref{ssect:acoustic}. The coupling to the
equations of state of linear piezoelectricity is discussed in subsection
\ref{ssect:elmech}. Finally, a simple postprocessing strategy to account for
the disturbance of the transducer signal by the experimental setup is discussed
in \ref{ssect:filter}.

\begin{lstlisting}[float,captionpos=b,keywordstyle=\bf, frame=lines, language=Python,basicstyle=\ttfamily\scriptsize, caption={Data structure
for the computational domain in {\tt{python}} module file {\tt{domain.py}}.},
label=code:domain]
import numpy as np

class Domain(object):
    def __init__(self, (zMin, zMax), Nz = 1000):
        (self.z,self.dz) = np.linspace(
            zMin,zMax,Nz,retstep=True,endpoint=False)
        self.mua = np.zeros(Nz)
        self.rho = np.zeros(Nz)
        self.v = np.ones(Nz)

    def _z2i(self, zVal):
        return int((zVal-self.z[0])/self.dz)
    
    def _idxSet(self, zMin, zMax):
        return np.logical_and(
            self.z >= zMin, self.z <= zMax)

    def setProperty(self, q, (zMin, zMax), qVal):
        q[self._idxSet(zMin,zMax)] = qVal
\end{lstlisting}

\subsection{Discretization of the computational domain}
\label{ssect:domain}

In principle, the above set of equations describes a continuous model. However,
so as to be able to numerically solve these equations for reasonable boundary
conditions and initial values, we need to consider a discrete realization of
the computational domain that holds local material properties such as $\rho(z)$
and $c(z)$. 
Later, we will consider the propagation of laser generated acoustic stress
waves within the medium. Therefore, the optical properties of the medium,
responsible for the absorption and scattering of laser light are of importance.
In our simplified $1$D model we will consider the limiting case of purely
absorbing media.  Therefore we here introduce a further local material
property, namely the absorption coefficient $\mu_a(z)$.
In order to set up a data structure that holds all these local material
properties, consider a discrete $z$-grid with constant mesh width, i.e.\ $z_i =
z_{\rm{min}} + i \Delta z$ for $i=0\ldots N-1$ and $\Delta z =
(z_{\rm{max}}-z_{\rm{min}})/N$, and let, e.g., $\rho_i$ refer to $\rho(z_i)$.
Then, a proper domain data structure might be implemented by the {\tt{python}} class
{\tt{Domain}} in code listing \ref{code:domain}. Subsequently we assume that the
respective code is available in the python module file {\tt{domain.py}}.

\subsection{Optical absorption of laser energy}
\label{ssect:opt}

In our model, the propagation of acoustic stress waves is triggered by an
initial distribution of acoustic stress $p(z,0)=p_0(z)$ for a medium at rest
$u(z,0)=0$. The nonzero initial pressure profile results from a thermoelastic
conversion of deposited laser energy to mechanical stress within the medium. 
In comparison to the typical timescale of mechanical response, the deposition
of laser energy can be considered instantaneous.
Below, the material parameter $\mu_a(z)$ accounts for the absorption of photons
within the medium. As a consequence, considering a light flux in $z$ direction,
a laser beam will experience a decrease of its incident fluence $f_0$ with
increasing depth $z$ following the Beer-Lambert decay law $f(z) = f_0
\exp\{-\int_{z_{\rm{min}}}^z \mu_a(z^\prime)~\mathrm{d}z^\prime\}$
\cite{Paltauf:2000}.  For such a purely absorbing medium, the amount of locally
absorbed laser energy $W(z)$ is related to the respective fluence decay through
$W(z) = - \mathrm{d}f(z)/\mathrm{d}z = \mu_a(z) f(z)$ \cite{Welch:1984}.
Finally, the efficiency of the conversion of deposited laser energy to acoustic
stress is governed by the Gr\"uneisen parameter $\Gamma$, by means of which
$p_0(z)=\Gamma\,W(z)$.  Note that the initial pressure pulses obtained in this
manner exhibit abrupt changes along the $z$ axis, signaling a sudden increase
or decrease of the absorption coefficient. Using these ``shockwave'' initial
conditions in the acoustic propagation algorithm implemented below might cause
numerical artefacts in the observables. In such a situation, if the absorption
coefficient is nonzero in a small range $z \in [z_p, z_p+\delta]$ only, a
remedy might be to use a simple Gaussian function with peak intensity $f_0$,
$1/e$ extension of $\delta$, and centered at $z=z_p + \delta/2$, instead.
Albeit this does not properly model the exponential attenuation of laser
fluence, it allows to study the principal distortion of pressure profiles upon
propagation.
Another possibility to circumvent the challenges such a shockwave might cause
for a finite-difference scheme is to adopt a high-precision finite-volume
procedure, see the discussion in sec.\ \ref{sec:S5}.

Once the computational domain is initialized and the optical properties
declared, a virtual laser beam can be propagated through the medium by means of
the function {\tt{absorbLaserBeam}} in code listing \ref{code:laserBeam}.
On input it expects three arguments, i.e.\ the details of the computational
domain {\tt{Glob}}, the Gr\"uneisen parameter {\tt{Gamma}} and the initial
laser fluence {\tt{f0}}. On output it yields a tuple consisting of the initial
pressure profile as a function of $z$ and the fluence of the transmitted part
of the beam. Further, an implementation of a simple Gaussian pressure profile 
as discussed above is provided by the function {\tt{deltaPulse}}.

\begin{lstlisting}[float,captionpos=b,keywordstyle=\bf, frame=lines, language=Python,basicstyle=\ttfamily\scriptsize, 
caption={Implementation of 
a function that yields an initial acoustic stress profile based on the optical
absorption of laser light by an attenuating medium, contained in {\tt{python}}
module file {\tt{opticalAbsorption.py}}.}, label=code:laserBeam]
import numpy as np

def deltaPulse(Glob,zRange,Gamma=1.,f0=1.):
    zMin, zMax = min(zRange), max(zRange) 
    z0, d = (zMax - zMin)/2, (zMin + zMax)/2
    return Gamma*f0*np.exp(-(Glob.z-z0)**2/d/d)

def absorbLaserBeam(Glob,Gamma=1.,f0=1.):
    fz = f0*np.exp(-np.cumsum(Glob.mua*Glob.dz))
    return Gamma*Glob.mua*fz, fz[-1] 
\end{lstlisting}

\subsection{Acoustic propagation via finite differences}
\label{ssect:acoustic}

The above hyperbolic set of partial differential equations might be discretized
via a staggered-grid leapfrog scheme \cite{NumRec:1992}.  Therefore, consider
an additional $t$-grid with constant mesh width $\Delta t$, i.e.\ let $t_n =
t_{\rm{min}} + n \Delta t$ for $n=0 \ldots M-1$ and $\Delta t =
(t_{\rm{max}}-t_{\rm{min}})/M$, and let, e.g., $p_i^n$ refer to $p(z_i,t_n)$.
Then, discretization yields the numerical approximations
\begin{subequations}
\begin{align}
  &[ K\, D_z u + D_t p]_i^n = 0,  \label{eq:finDiffEq1}\\ 
  &[\rho\, D_t u + D_z p]_i^n = 0,  \label{eq:finDiffEq2} 
\end{align}
\end{subequations}
of Eqs.\ (\ref{eq:linEl1}) and (\ref{eq:linEl2}), respectively, wherein the
centered half-step grid-derivatives for the exemplary variable $u$ read 
\begin{subequations}
\begin{align}
 &[D_t u]_i^n = \frac{1}{\Delta t}\Big(u_i^{n+\frac{1}{2}} - u_i^{n-\frac{1}{2}}\Big), \label{eq:Dt}\\
 &[D_z u]_i^n  = \frac{1}{\Delta z}\Big(u_{i+\frac{1}{2}}^n - u_{i-\frac{1}{2}}^n \Big). \label{eq:Dz}
\end{align}
\end{subequations}
For convenience, shifting the velocity field to the intermediate
coordinates, i.e.\ letting $n \to n+1/2$ in Eq.\ (\ref{eq:finDiffEq1})
and $i \to i+1/2$ in Eq.\ (\ref{eq:finDiffEq2}), and using the harmonic 
mean $\rho_{i+\frac{1}{2}}^{-1} = (\rho_i^{-1} + \rho_{i+1}^{-1})/2$,
we obtain the evolution equations
\begin{subequations}
\begin{align}
&u_{i+\frac{1}{2}}^{n+\frac{1}{2}} ~=~ u_{i+\frac{1}{2}}^{n-\frac{1}{2}} ~-~ \rho_{i+\frac{1}{2}}^{-1} \frac{\Delta t}{\Delta z} \Big(p_{i+1}^n - p_i^n\Big), \label{eq:evolve_u}\\
&p_{i}^{n+1} ~=~ p_{i}^{n} ~-~ K_{i} \frac{\Delta t}{\Delta z} \Big(u_{i+\frac{1}{2}}^{n+\frac{1}{2}} - u_{i-\frac{1}{2}}^{n+\frac{1}{2}}\Big), \label{eq:evolve_p}
\end{align}
\end{subequations}
for the acoustic velocity and excess pressure fields.
On either end of the computational domain, the boundary conditions implement
pressure-release boundaries, i.e.\ we impose $p_0^n=0$ and $p_{N-1}^n$ at 
each timestep, defining a free surface.
A vectorized {\tt python} implementation of the above formulae is shown 
in code listing \ref{code:fdScheme}.
The first $4$-tuple of parameters specify the initial values for the excess
pressure ({\tt{p0}}) and the details of the computational domain, i.e.\ sonic
velocity ({\tt{v}}), density ({\tt{rho}}), and mesh width ({\tt{dz}}). The
parameter {\tt measure} refers to a function that facilitates the monitoring of
relevant observables.  This callee will be defined later on.  However, note
that it receives four arguments that specify the current time step ({\tt{n}}),
the time increment ({\tt{dt}}), the full velocity field ({\tt{u}}) and excess
pressure ({\tt{p}}).  Finally, {\tt{Nt}} signifies the overall number of
simulation time steps until termination of the propagation process.

\begin{lstlisting}[float,captionpos=b,keywordstyle=\bf, frame=lines, language=Python,basicstyle=\ttfamily\scriptsize, 
caption={Implementation of
the staggered leapfrog finite difference equations for propagating the velocity
and excess pressure fields, contained in {\tt{python}} module file
{\tt{stressWavePropagation1DSLDE.py}}.}, label=code:fdScheme]
import numpy as np

def propagateStressWaveSLDE((p0,v,rho,dz),measure,Nt):
    dt = 0.3*dz/max(v)
    C = dt/dz
    u1 = np.zeros(p0.size-1)
    u = np.zeros(p0.size-1)
    p1 = np.copy(p0)
    p = np.zeros(p0.size)
    ri = (1./rho[:-1] + 1./rho[1:])/2 
    K = rho*v*v

    measure(0, dt, u, p)

    for n in range(1,Nt): 
        # advance material velocity
        u[:] = u1[:] - ri[:]*C*(p1[1:]-p1[:-1])
        # advance stress
        p[1:-1] = p1[1:-1] - K[1:-1]*C*(u[1:]-u[:-1]) 
        # enforce BCs on stress profile
        p[0] = 0.; p[-1]=0.
        # measurement at timestep n
        measure(n, dt, u, p)
        # advance timestep
        u1[:], p1[:] = u[:], p[:] 

    return p1
\end{lstlisting}

\subsection{Realizing the piezoelectric coupling}
\label{ssect:elmech}

The implementation of a piezoelectric sensing layer requires the
electromechanical coupling of the equations of linear elasticity to the state
equations of $1$D piezoelectricity.  Here we consider a piezoelectric layer as
a stress sensing device, relying on the direct piezoelectric effect in which a
mechanical load, i.e.\ stress within the material, is converted to an electric
field. If electrodes are connected to the opposing layer surfaces, the
resulting potential difference across the layer might be measured.  The
theoretical framework in which the piezoelectric response might be described
depends on mechanical and electrical boundary conditions (BCs).  Depending on
the subtleties of the experimental setup that needs to be modeled, one
distinguishes a mechanically free setup (i.e.\ at constant stress), labelled
``T'', and a mechanically clamped setup (i.e.\ at constant strain), labelled
``S'', as well as an electrical short-circuit setup (i.e.\ at constant
electrical field), labelled ``E'', and an electrical open-circuit setup (i.e.\
at constant electrical displacement), labelled ``D''.  Here, we consider the
idealized case of a mechanically free setup with an open-circuit. 
The respective $1$D constituting relations read
\begin{subequations}
\begin{align}
&S = s^{D}\,T + g\,D, \label{eq:piezoEq1} \\
&D = d\, T + \epsilon^T\,E,\label{eq:piezoEq2}
\end{align}
\end{subequations}
wherein $E$, $D$, $S$ and $T$ refer to the electric field, field displacement,
strain and stress, respectively. Further, $s^{D}$ denotes the mechanical
compliance, $d$ refers to the piezoelectric strain constant and $\epsilon^{T}$
signifies the dielectric coefficient
at constant stress.
Note that under the assumption of a vanishing charge density in the unstressed
state of the layer, the field displacement $D$ is zero. Further, considering
the displacement gradient formulation of the strain, i.e.\ $S(z,t) = \partial_z
r(z,t)$ and thus $\partial_t S(z,t) = \partial_z u(z,t)$, 
allows to cast Eqs.\ (\ref{eq:piezoEq1}) and (\ref{eq:piezoEq2}) into
the computationally convenient form
\begin{equation}
\partial_t E(z,t) + h \, \partial_z u(z,t) = 0 \label{eq:piezoEl_cont}
\end{equation}
of the direct piezoelectric effect, wherein $h = d/(\epsilon^{T} s^{D})$.
Again, using centered half-step grid derivatives as above, this can be 
approximated by the finite difference equation 
\begin{equation}
[D_t E + h \, D_z u]_i^n = 0, \label{eq:piezoEl_fd}
\end{equation}
providing an evolution equation for the electric field in the piezoelectric
layer in the form
\begin{equation}
E_{i}^{n+1} ~=~ E_{i}^{n} ~-~ h\frac{\Delta t}{\Delta z} \Big(u_{i+\frac{1}{2}}^{n+\frac{1}{2}} - u_{i-\frac{1}{2}}^{n+\frac{1}{2}}\Big). \label{eq:evolve_E}
\end{equation}
Finally, the potential difference $U(t)=\phi(z_{+})-\phi(z_{-})$ between the
opposing layer surfaces (located at $z_{-}$ and $z_{+}$) can be obtained from
$E(z,t) = - \partial_z \phi(z,t)$ by numerical quadrature, e.g.\ using a
trapezoidal rule. 
Note that, considering the constituting equation of the direct piezoelectric
effect, i.e.\ Eq.\ (\ref{eq:piezoEq2}) together with the $1$D stress-pressure
relation $T(z,t)=-p(z,t)$ of hydrostatic compressions \cite{Landau:1975}, the
potential difference $U(t)$ between the opposing layer surfaces (located at a
distance $\ell=z_{+}-z_{-}$) can be found to be
\begin{align}
U(t) = - \frac{d \ell}{\epsilon^T} \Big( \frac{1}{\ell} \int_{z_{-}}^{z_{+}} p(z,t)~\mathrm{d}z \Big).\label{eq:avPressure}
\end{align}
Thus, the potential difference is simply proportional to the average pressure
within the piezoelectric sensing layer, i.e.\ $U(t)\propto \bar{p}(t)$.  As
pointed out earlier, the above finite difference scheme was derived for the
idealized setup considering the mechanically free BCs.  Simulations for a
mechanically clamped setup can easily be done by setting $h = e/\epsilon^S$ in
Eq.\ (\ref{eq:evolve_E}), wherein $e$ refers to the piezoelectric stress
constant and $\epsilon^{S}$ signifies the dielectric constant at fixed stress.
This follows from considering the piezoelectric constituting equations for BCs
``S'' and ``D''.  

A proper transducer data structure that implements the finite difference
approximation to the linear piezoelectric constituting equations is given by
the {\tt{python}} class {\tt{PiezoTransducer}} shown in code listing
\ref{code:transducer}.  An instance of the transducer class expects three
arguments on input: an instance {\tt{Glob}} of the discrete computational
domain, a tuple {\tt{(zMin, zMax)}} specifying the surface locations of the
sensing layer and an optional argument {\tt{h}} that combines the piezoelectric
and mechanical system parameters as discussed above.  Note that the class
provides a method called {\tt{measure}} that can be passed as callback function
to the {\tt propagateStressWaveSLDE} routine, assuming the role of the function
{\tt measure} in \mbox{code listing \ref{code:fdScheme}.}

\begin{lstlisting}[float,captionpos=b, keywordstyle=\bf, frame=lines, language=Python,basicstyle=\ttfamily\scriptsize, 
caption={Implementation of a
piezoelectric transducer as piezoelectric sensing layer, contained in {\tt{python}}
module file {\tt{detector.py}}.}, label=code:transducer]
import sys
import numpy as np

class PiezoTransducer(object):

    def __init__(self,Glob,(zMin,zMax),h=1.):
        self.dz = Glob.dz
        self.zIdMin = max(1,Glob._z2i(zMin))
        self.zIdMax = Glob._z2i(zMax)
        self.E = np.zeros(self.zIdMax-self.zIdMin)
        self.h = h
        self.t = []
        self.U = []

    def measure(self,n,dt,u,tau):
        C = dt/self.dz
        E0 = self.E 
        h = self.h
        zL, zH = self.zIdMin, self.zIdMax

        # evolve electric field within transducer
        self.E[:] = E0[:] - h*C*(u[zL-1:zH-1]-u[zL:zH])
        # potential difference across transducer 
        dU = -np.trapz(self.E,dx=self.dz)

        self.t.append(n*dt)
        self.U.append(dU)

\end{lstlisting}

\subsection{Signal postprocessing in the time domain}
\label{ssect:filter}

Note that the experimental setup used to amplify and postprocess the  
piezoelectric response might have an impact on the shape of the signal itself.  In
order to mimic such a disturbance one might pursue one out of several
postprocessing strategies, based on digitally filtering the detected signal in
the time- or frequency domain \cite{NumRec:1992}.  Such a filter takes an input
signal $s(t)$, i.e.\ a time-series of input points, and yields a modified
output signal $s^\prime(t)$ subject to several physical constraints.  E.g., a
time-domain filter might work online, processing input points as they are
recorded, or offline, as a convenient postprocessing device.  While the former
working mode naturally ensures causality since it prohibits the filter to
access input points that are out of time, yet, the latter working mode allows
also for a more general behavior wherein a certain input point might depend on
earlier as well as later input points. Here, for convenience, we opt for an
offline filter that works in the time domain and obeys causality.  Without
elaborating on the subtleties of the experimental setup used in our laboratory
experiments, we here assume that the experimental setup has the characteristics
of a first order low pass filter, suppressing high frequency features of the
input signal. This choice is purely phenomenologic since the effect of such a
filter is consistent with our observations.

In general, a low-pass filter is characterized by time constant $\tau$,
defining its ``cutoff frequency'' $\omega_{\rm c}=\tau^{-1}$. Signal
features with a frequency $\omega > \omega_{\rm c}$ appear suppressed. If the
input signal is sampled at constant time increment $\Delta t$, one might also
define the filter parameter $a = \Delta t/(\tau+\Delta t)$, with $0\leq a \leq
1$, to characterize its performance.  Denoting input and output signals as $s$
and $s^\prime$, respectively, the effect of a simple first order low-pass
filter might be cast into the recurrence relation 
\begin{equation}
s^\prime_i = a s_i + (1-a) s^\prime_{i-1}. \label{eq:lowPass}
\end{equation}

A data structure that implements such a low-pass filter is shown in code
listing \ref{code:filter}. Therein an instance of the time domain filter class
{\tt{Filter}} expects two arguments on input: the ordered sequence of
time samples {\tt{t}} and the input signal {\tt{s0}}.  The low-pass is provided
as method {\tt{lowPass}}. As an argument, the method takes the characteristic
smoothing parameter {\tt{a}} of the filter and yields an output signal that
contains the distortion of the input signal by the low pass.  The filter
described above can be used to postprocess the detector signals recorded via
the piezoelectric transducer implemented in code listing \ref{code:transducer}.

\begin{lstlisting}[float,captionpos=b, keywordstyle=\bf, frame=lines, language=Python,basicstyle=\ttfamily\scriptsize, 
caption={Implementation of a low-pass filter, contained in
{\tt{python}} module file {\tt{timeDomainFilter.py}}.}, label=code:filter]
import numpy as np

class Filter(object):

    def __init__(self,t,s0):
        self.t = np.asarray(t)
        self.dt = self.t[1]-self.t[0]
        self.s0 = np.asarray(s0)

    def lowPass(self,a):
        s = np.zeros(self.s0.size)
        s[0] = self.s0[0]
        for i in range(1,s.size):
            s[i] = a*self.s0[i] + (1.-a)*s[i-1]  
        return s

\end{lstlisting}

\section{Limitations of the one-dimensional approach}
\label{sec:S3}

Since the theoretical approach implemented in sect.\ \ref{sec:S2} is tailored
to a particular physical scenario, there are limitations that restrict its
range of application. Below, without any claim on completeness, we address some 
of these limitations.

In principle, considering a nonscattering $3$D setup wherein $\vec{r}=(x,y,z)$,
the deposition of energy within the medium is governed by laser beam parameters
such as the incident laser fluence at maximal intensity $f_0$, the transverse
beam profile $f(x,y)$ and the temporal intensity profile $I(t)$ of the laser
pulse, as well as the absorption coefficient $\mu_a(x,y,z)$ within the medium.
Two prerequisites allow for an effectively $1$D
treatment of the problem of optical absorption: (i) the absorption coefficient
depends only on the depth coordinate $z$, i.e.\ $\mu_a(x,y,z)\equiv \mu_a(z)$,
thus realizing a layered medium with translational invariance in the
$(x,y)$-plane, and, (ii) the transverse beam profile is wide enough to ensure 
initially plane acoustic wavefronts.  
This can be accomplished by requiring the penetration depth $\ell$ of the laser
into the medium, defined via $1\equiv \int_{z_{\rm{min}}}^{\ell}
\mu_a(z^\prime)~\mathrm{d}z^\prime$, to be smaller than the characteristic
lengthscale, say, e.g., the beam diameter $d$, that defines the extension of the
laser spot on the medium (assuming a constant laser intensity within the spot).
Then, the validity of the $1$D approach is limited by the onset of diffraction 
during the propagation of the initial pressure profile along the $z$ axis.

In our numerical approach, the propagation of stress waves is accomplished by a
finite-difference approximation of the equations of linear elasticity, i.e.\
Eqs.\ (\ref{eq:linEl1}) and (\ref{eq:linEl2}), under the assumption of initial
plane acoustic waves. While this approach seems sufficient for our particular
application, see the results reported in sect.\ \ref{sec:S4}, there are several
options to extend the numerical procedure.  E.g., if radial inertia following
the optical absorption of laser energy might not be neglected, Love's modified
wave equation might be used to include the effect of dispersive waves during
the propagation process \cite{Percival:1969}.  If elastic media with losses,
as, e.g.\ tissue, are considered, a Kelvin-Voigt model wave equation might be
used to account for viscoelastic effects \cite{Szabo:2004}.  Further, note that
if the assumption of initial plane acoustic waves and stress wave detection in
the acoustic nearfield are not satisfied, a more complete $3$D description of
the problem that naturally accounts for the effect of acoustic diffraction is
needed.  Also note that since we study the propagation of stress waves prior to
thermal equilibration of the source volume, which occurs on larger timescales,
we neglect diffusion of heat within the computational domain. 

For our numerical experiments we consider a PVDF (polyvinylidene fluoride)
polymer as piezoelectric sensing layer.  Note that PVDF exhibits not only
piezoelectric but also pyroelectric properties \cite{Ueberschlag:2001}.  Thus,
if the temperature of the PVDF layer is expected to increase notably within the
duration of the measurement process, pyroelectric effects might be expected in
addition to the direct piezoelectric effect discussed in subsect.\
\ref{ssect:elmech}. However, in the particular source volume configuration 
studied in the presented article, the initial stress pulse is separated from 
the PVDF layer by an approx.\ $0.5\,{\rm{mm}}$ wide PMMA layer. Since the
thermal diffusivity of PMMA amounts to $\alpha = 0.1054\,{\rm mm^2/s}$ (at a
temperature of $T=25^\circ$, see Ref.\ \cite{NunesDosSantos:2005}), the 
approximate thermal relaxation time for thermal equilibration over the 
former distance is $t_{\rm{r}}=9.49\,{\rm{s}}$, exceeding the observation 
time by several orders of magnitude. Hence, as pointed out above, we neglect
heat diffusion and thus also pyroelectric effects in the PVDF layer.

\section{Results and Discussion}
\label{sec:S4}

\begin{figure}[t!]
\begin{center}
\includegraphics[width=1.0\linewidth]{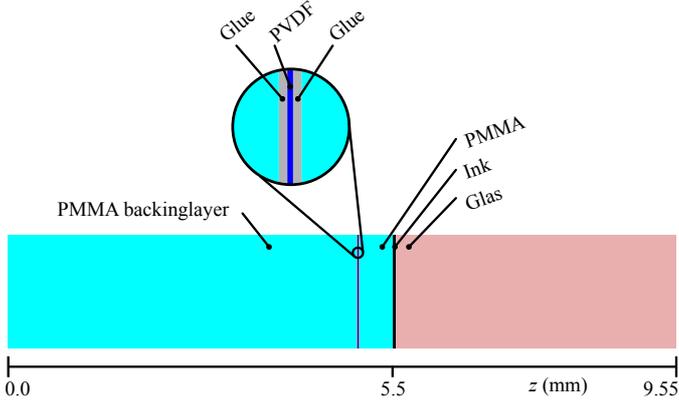}
\end{center}
\caption{
(Color online) Configuration of the $1$D computational domain used in the
numerical experiments.  Layers from left to right: PMMA, Glue, PVDF, Glue,
PMMA, Ink, Glass.  The zoom-in gives an enlarged view on the piezoelectric
transducer, ``sandwiched'' in between glue and PMMA layers.  The
configurational properties of the individual layers are listed in table Tab.\
\ref{tab:simPars}.  \label{fig:cfg}}
\end{figure}  

The numerical approach detailed in sect.\ \ref{sec:S2} allows to implement a
piezoelectric sensing sensing layer with or without acoustic backing layer.
Both setups allow for notably different signals that might be expected.  Here,
in order to perform custom numerical simulations for an experimental setup we
seek to match, we consider a PVDF layer glued (using Norland Optical Adhesive
``NOA85V'') in between two layers of PMMA (polymethylmethacrylat), serving as
backing- and frontlayer.  As illustrated in Fig.\ \ref{fig:cfg}, we assume that
the deposition of laser energy in the computational domain occurs in a thin
layer of black ink in between the PMMA frontlayer and a closing plate of glass.
The configuration of the domain, i.e.\ the thickness of all layers along with
their assumed optic and acoustic properties, is listed in Tab.\
\ref{tab:simPars}.  Further details regarding the experimental setup are
discussed elsewhere \cite{Blumenroether:2017}.

\begin{table}[b!]
\caption{\label{tab:simPars} 
Configuration and simulation parameters of the computational domain. From left
to right: layer material, $[z_{-},z_{+}]$-range of layer (in ${\rm{mm}}$),
density $\rho$ (in ${\rm{mg/mm^{3}}}$), wave velocity $c$ (in ${\rm{mm/ms}}$),
absorption coefficient $\mu_a$ (in ${\rm{mm}^{-1}}$) and layer impedance $Z$
(in ${\rm{mg/(ms\, mm^{2})}}=10^2 {\rm{Rayl}}$).
}
\begin{tabular}[c]{llllll}
\hline
Material & $[z_{-},z_{+}]$ & $\rho$   & $c$ & $\mu_a$ & $Z$  \\
\hline \hline 
PMMA  & $[0.000, 4.980]$ & 1.18 & 2.77 & 0.0   & 3.3   \\
Glue  & $[4.980, 4.995]$ & 1.00 & 2.50 & 0.0   & 2.5    \\
PVDF  & $[4.955, 5.005]$ & 1.78 & 2.25 & 0.0   & 4.0    \\
Glue  & $[5.005, 5.020]$ & 1.00 & 2.50 & 0.0   & 2.5    \\
PMMA  & $[5.020, 5.500]$ & 1.18 & 2.77 & 0.0   & 3.3    \\
Ink   & $[5.500, 5.550]$ & 1.00 & 1.80 & 100.0 & 1.8    \\
Glass & $[5.550, 9.550]$ & 2.23 & 5.60 & 0.0   & 12.5   \\
\hline
\end{tabular}
\end{table}

\begin{figure}[t!]
\begin{center}
\includegraphics[width=1.0\linewidth]{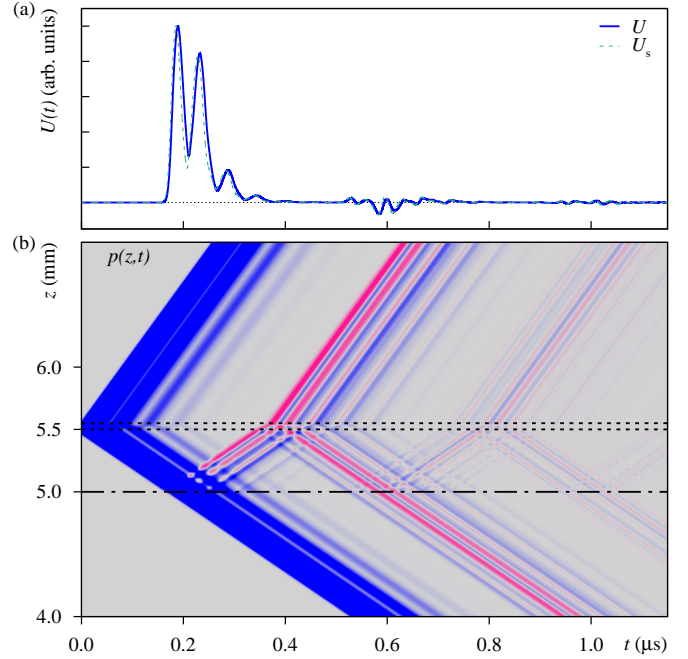}
\end{center}
\caption{
(Color online) (a) Response $U(t)$ of the piezoelectric sensing layer to
trespassing acoustic stress waves, auxiliary signal $U_{\rm s}(t)$ assumed to
be proportional to the local pressure at the PVDF surface facing the PMMA
frontlayer, and, (b) space-time plot $p(z,t)$ of the $1$D acoustic stress
profile within the $[4\,{\rm{mm}},7\,{\rm{mm}}]\times[0\,{\rm{\mu
s}},1.1\,{\rm{\mu s}}]$ region of the computational domain.
Compression (rarefaction) phases are colored blue (red). 
The dashed lines at $z=5.5\,{\rm{mm}}$ and $5.55\,{\rm{mm}}$
 mark the boundaries of the ink layer and the dash-dotted line at 
$z=5.005\,{\rm{mm}}$ indicates the front surface of the PVDF sensing layer.
\label{fig:pzt}}
\end{figure}

\paragraph{Analysis of the transducer response}
Starting from an initial stress profile, obtained following the OA signal
generation procedure outlined in subsect.\ \ref{ssect:opt} and evolved using the
staggered-grid leapfrog scheme presented in subsect.\ \ref{ssect:acoustic},
Fig.\ \ref{fig:pzt}(a) shows the time sequence of the potential difference
$U(t)$ between the boundaries of the PVDF layer.  As argued in subsect.\
\ref{ssect:elmech}, the resulting Voltage signal is proportional to the average
pressure in the piezoelectric sensing layer.  Assuming that the intrinsic
lengthscale $\lambda_{\rm{ac}} = 2 \ell$ (wherein $\ell$ is the penetration
depth of the laser defined in sect.\ \ref{sec:S3}) that might be used to define
the acoustical wavelength in the given setup exceeds the thickness of the
piezoelectric layer, the Voltage $U_{\rm{s}}(t)$ caused by the
trespassing stress wave is simply proportional to the local pressure at, say,
the surface facing the PMMA frontlayer, i.e.\ $U_{\rm{s}}(t)\propto
p(z^{\rm{PVDF}}_{+},t)$ \cite{Schoeffmann:1988,Jaeger:2005} (in our numerical
experiment $z^{\rm{PVDF}}_{+}=5.005\,{\rm{mm}}$). If the width of the PVDF
layer exceeds $\lambda_{\rm{ac}}$ notably, both Voltage signals are expected to
differ significantly \cite{Schoeffmann:1988}. We verified this behavior by
means of further numerical experiments based on a simpler design of the
computational domain (not shown).  In this particular application we assume the
width of the PVDF layer $\Delta z_{\rm{PVDF}}=10\,{\rm{\mu m}}$ and the
acoustical wavelength $\lambda_{\rm{ac}}=20\,{\rm{\mu m}}$ to be approximately
of the same extend. As evident from Fig.\ \ref{fig:pzt}(a), both signals are
(still) in reasonable agreement as expected for a ``thin'' sensing layer and a
``wide'' trespassing stress wave.

An implementation of the computational domain using the simulation parameters
listed in Tab.\ \ref{tab:simPars} is shown in code listing \ref{code:numExp}.
Provided that the imported modules are in the local searchpath, the necessary
statements to produce a plot similar to Fig.\ \ref{fig:pzt}(a) requires merely
$48$ lines of {\tt{python}} code.

Note that the time sequence of the transducer response $U(t)$ features several
groups of features: the first group reaches the PVDF layer in the interval
$t\in (0.15,0.45)\,{\rm{\mu s}}$, resulting from the initial stress wave
passing the Ink-PMMA (reflection coefficient $C_r({\rm{Ink}},{\rm{PMMA}})
\approx 0.29$) and PMMA-PVDF inter-layer boundaries also after multiple
reflections within the Ink layer, enclosed bye PMMA and Glass
($C_r({\rm{Ink}},{\rm{Glass}}) \approx 0.74$). Since both relevant reflection
coefficients are positive, the first group of features is a train of
compression peaks. The first reflection with a negative reflection coefficient
occurs at the PMMA-Glue inter-layer crossing ($C_r({\rm{PMMA}},{\rm{Glue}})
\approx -0.13$), triggering a rarefaction wave that travels back towards the
PMMA-Ink interface where its sign changes again due to the apparent coefficient
$C_r({\rm{PMMA}},{\rm{Ink}}) \approx -0.29$, approaching the sensing layer,
once again, as compression wave. However, due to multiple inter-layer
reflections that involve coefficients $C_r<0$, rarefaction waves eventually
trespass the PVDF layer, see, e.g.\ the second group of features reaching the
PVDF layer in the interval $t\in (0.50,0.75)\,{\rm{\mu s}}$. The sequence of
sign changes in the $1$D pressure profile $p(z,t)$ can be disentangled best
using the space-time plot shown in Fig.\ \ref{fig:pzt}(b).

\begin{lstlisting}[float,captionpos=b, keywordstyle=\bf, frame=lines, language=Python,basicstyle=\ttfamily\scriptsize, 
caption={Implementation of the examplary application discussed in the sec.\
\ref{sec:S3}, contained in {\tt{python}} script file
{\tt{main\_InkOnGlass\_Fig2a.py}}.  Given that the imported modules are in the
local searchpath, the script produces output similar to Fig.\,
\ref{fig:pzt}(a)}, label=code:numExp]
import matplotlib.pyplot as plt
from domain import *
from detector import *
from opticalAbsorption import *
from stressWavePropagation1DSLDE import *

def main():
     # SIMULATION PARAMETERS 
     Nt = 50000
     zMax, Nz = 9.55, 8600        

     # SET LAYER STRUCTURE OF MEDIUM
     # LayerNo: ((zMin, zMax), (c, rho, mua))
     # Units: [c]=mm/mus, [rho]=mg/mm, [mua]=1/mm
     # (1) PMMA, (2) Glue, (3) PVDF, (4) Ink, (5) Glass
     layers = {
         1: ((0.000, 5.500), (2.77, 1.18, 0.000)), 
         2: ((4.980, 5.020), (2.50, 1.00, 0.000)), 
         3: ((4.995, 5.005), (2.25, 1.78, 0.000)), 
         4: ((5.500, 5.550), (1.80, 1.00, 100.0)), 
         5: ((5.550, 9.550), (5.60, 2.23, 0.000))  
     }

     # INSTANCE OF COMPUTATIONAL DOMAIN
     Glob = Domain((0.,zMax), Nz)
     # INSTANCE OF DETECOR FOR MONITORING OBSERVABLES
     Det  = PiezoTransducer(Glob, (4.995,5.005))

     # ASSIGN LAYER PROPERTIES TO COMPUTATIONAL DOMAIN
     for no, (zR, (c, rho, mu)) in \
             sorted(layers.iteritems()):
         Glob.setProperty(Glob.v, zR, c) 
         Glob.setProperty(Glob.rho, zR, rho)                 
         Glob.setProperty(Glob.mua, zR, mu)
     
     # OPTICAL ABSORPTION
     p0 = absorbLaserBeam(Glob)
     # ACOUSTIC PROPAGATION
     p = propagateStressWaveSLDE(
         (p0,Glob.v,Glob.rho,Glob.dz), Det.measure,Nt)

     # DISPLAY RESULTS 
     plt.plot(Det.t,Det.p/max(Det.p))
     plt.xlabel('t (mu s)')
     plt.ylabel('U(t) (a.u.)')
     plt.show()

main()
\end{lstlisting}

\begin{figure}[t!]
\begin{center}
\includegraphics[width=1.0\linewidth]{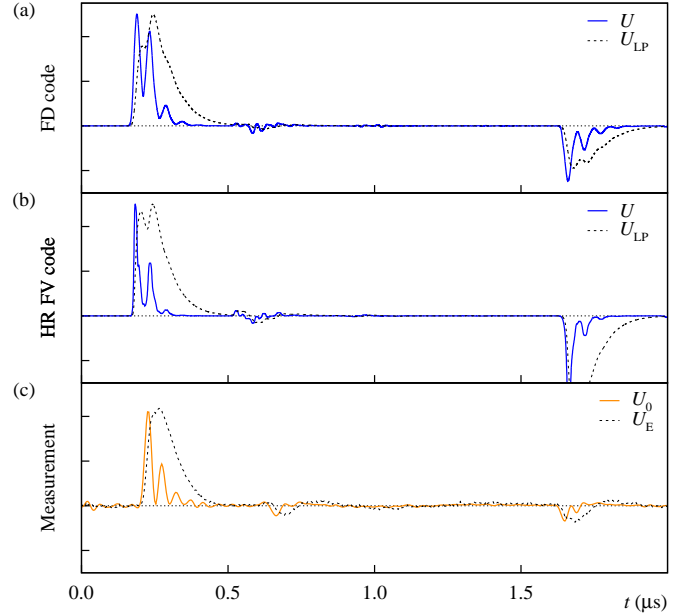}
\end{center}
\caption{
(Color online) Transducer response and distortion of the signal by the ``black
box'' experimental setup.  (a) Results of the numerical experiments in terms
of the $1$D finite-difference (FD) wherein
$U(t)$ refers to the potential difference between the boundaries of the
piezoelectric sensing layer, and where $U_{\rm{LP}}(t)$ accounts for a possible
distortion of the signal using a low-pass filter characteristic as discussed in
the text,
(b) results of a reference simulation using a high-resolution finite-volume (HR FV)
solver (see text for details), and,
(c) measured transducer response $U_{\rm{E}}$, including the
distortion of the underlying signal by the experimental setup, and signal
$U_{\rm{0}}$, obtained after correcting for the signal distortion via the
transducers transfer-function.
\label{fig:TvsE}}
\end{figure}

Assuming that the impact of the experimental setup on the transducer signal can
be described by means of a low-pass filter with time constant $\tau \approx
0.05\,{\rm{\mu s}}$ effectively smoothes the input transducer signal and yields
an output signal as shown in Fig.\ \ref{fig:TvsE}(a). Therein, the grouped
features of the input signal are partially lost due to the attenuation of
frequencies above the cutoff $\omega \approx 20 \times 10^6\,{\rm{rad/s}}$ (i.e.\ $f =
\omega/ 2 \pi\approx 3.2\,{\rm{MHz}}$) in the ultrasonic frequency range.

This completes the interpretation and analysis of transducer signals obtained
for the computational domain shown in Fig.\ \ref{fig:cfg}. The comparison of
the numerical results to experimental data follows below.

\paragraph{Comparison to experimental results}

In Fig.\ \ref{fig:TvsE} we compare the numerical simulations to experimental
data, obtained from measurements of OA stress waves via a PVDF transducer 
using a configuration similar to Fig.\ \ref{fig:cfg}. 
The figure allows to compare the predicted transducer response (including the
assumed low-pass distortion due to the experimental setup), to the measured
response, cf.\ the black dashed lines in Figs.\ \ref{fig:TvsE}(a),(c).  The
inferred transducer response, obtained by correcting for the transducer
transfer-function in the experiment \cite{Blumenroether:2017}, also compares
well to the calculated response of the PVDF layer, cf.\ the solid lines in
Figs.\ \ref{fig:TvsE}(a),(c).  As evident from the figure, the sequence of
compression and rarefaction pulses observed in the numerical and laboratory
experiments are in excellent agreement. 
Note that the group of rarefaction pulse signal features in the range $t \in
(1.6,1.9)\,{\rm{\mu s}}$, resulting from a reflection of the acoustic wave on
the far end of the glass layer, has a notably lower amplitude in the
experiment. This might be due to the circumstance that the stress wave needs to
traverse the liquid-solid ink-glass and ink-PMMA inter-layer boundaries
possibly several times. Thus, loss effects that are relevant in crossing such a
boundary, not accounted for in our approach, might be necessary to explain
this observation.

\paragraph{Verification using a finite-volume approach}

Modeling wave propagation in nonconservative hyperbolic systems, such as linear
elastic wave propagation in varying heterogeneous media governed by Eqs.\
(\ref{eq:linEl1})-(\ref{eq:linEl2}), render a challenge for finite-difference
methdos. Alternatively one might consider high-resolution finite-volume methods
originally developed for nonlinear problems \cite{Fogarty:1999,LeVeque:1997}.
Albeit computationally more expensive than the finite difference approach
illustrated in sect.\ \ref{sec:S2}, they allow for a solution of problem
instances that are not amenable to finite-difference methods, such as, e.g.\
shock-wave propagation in highly discontinuous nonlinear media
\cite{Berezovski:2006}. For numerical redundancy we verified the results, i.e.\
location of signal features and ordering of compression and rarefaction pulses,
obtained using our (simple) finite-difference approach via an independent
(elaborate) finite-volume implementation in terms of an acoustic $1$D Riemann
solver, provided by the {\tt{PyClaw}} {\tt{CLAWPACK}} tool
\cite{Ketcheson:2012,Clawpack:2017}, cf.\ Figs.\ \ref{fig:TvsE}(a) and (b).

\section{Conclusions}
\label{sec:S5}

In the presented article we considered the problem of optoacoustic generation
and propagation of stress waves within a piecewise homogeneous material, with
focus on their detection using a piezoelectric sensing layer.  The specific
objective of this study was to implement a simple numerical model that
facilitates the simulation of the transducer response in the acoustic nearfield
in terms of an effectively $1$D finite difference approach, and to complement
experimental results via numerical simulations.  Comparing numerical
simulations and experimental data obtained for a certain layered setup of the
source volume, we found that the portrayed numerical approach accurately
predict its elastic and piezoelectric response. In turn, the modeling of the
piezoelectric transducer response in the presented approach proved useful for
interpreting measured transducer signals and for verifying the assumed
transfer-function of the employed transducer.

Since the non-availability of code impedes transparency and reproducability of
results in scientific publications \cite{Barnes:2010,Ince:2012,Sandve:2013} we
considered it useful to make the concise research-code for the presented study,
including but not limited to the code listings \ref{code:domain} through
\ref{code:numExp} along with all scripts needed to reproduce all figures,
publicly available \cite{Melchert_gitHub_1DFD:2017}.

\section*{Acknowledgments}
This research work received funding from the VolkswagenStiftung within the
``Nieders\"achsisches Vorab'' program in the framework of the project ``Hybrid
Numerical Optics''  (HYMNOS; Grant ZN 3061). The software was developed and
tested under OS X Yosemite (Version: 10.10.3) on a MacBook Air featuring a
1.7GHz Itel Core i5 processor and 4 GB DDR3 using {\tt{python}}
\cite{Oliphant:2007} version $2.7.6$ and {\tt{numpy}} version $1.8.0{\rm{rc}}1$
\cite{Jones:2001}. 

\bibliography{commentsBibfile,masterBibfile}

\end{document}